\documentstyle[aps,eqsecnum,prbbib]{revtex}  
\input epsf
\begin{document}
\draft
 \title{Tunneling spectroscopy for 
ferromagnet/superconductor junctions}
\author{Igor \v{Z}uti\'c\cite{igor}}
\address{Department of Physics, 
University of Maryland, College Park, Maryland 20742} 
\author{Oriol T. Valls\cite{oriol}}
\address{Department of Physics and Minnesota Supercomputer 
Institute, University of Minnesota,
Minneapolis, Minnesota 55455}
\date{\today}
\maketitle
\begin{abstract}
In tunneling spectroscopy studies of ferromagnet/superconductor
(F/S) junctions, the effects of spin polarization,
Fermi wavevector mismatch (FWM) between the F and S regions, and interfacial 
resistance play a crucial role. We study the low bias conductance 
spectrum of these junctions, governed by  Andreev reflection at
the F/S interface.  
We consider both  $d$- and $s$-wave superconductors
as well as mixed states of the $d+is$ form.
We present results for a range  of values of the relevant parameters
and find that a rich variety of features appears, depending
on pairing state and other conditions.
We show that in the presence of FWM, spin 
polarization can enhance
Andreev reflection and give rise to a zero bias conductance peak for
an $s$-wave superconductor.
\end{abstract}
\pacs{74.80.Fp, 74.50+r, 74.72-h}
\section{Introduction}
\label{intro}
The  development and refinement in recent years of new techniques in materials 
growth has made it possible 
to fabricate superconducting heterostructures with various materials and
high quality interfaces. These advances, coupled
with the continuing intense level of activity in the study of the nature of
high temperature\cite{agl,dj,har} and other exotic 
superconductors,\cite{upt,mae} 
has led to renewed interest in 
tunneling spectroscopy.

It has been demonstrated\cite{har,hu,tan} that this
technique yields information about 
both the magnitude and  the phase of the superconducting pair potential (PP). 
This implies that the method  can 
provide a systematic way to distinguish among various proposed PP 
candidates, including both spin singlet and spin triplet 
pairing states.\cite{tan3,sig}
For example, it has been argued that the observed zero bias conductance 
peak\cite{hu,tan,xu,wei,alf} 
(ZBCP),
attributed to mid-gap surface states, is an indication of 
unconventional superconductivity with a sign change of the PP,
as it occurs in pairing with a 
$d_{x^2-y^2}$-wave symmetry. Furthermore, the splitting of the ZBCP and 
the forming 
of a finite bias peak  (FBCP) in the conductance spectrum  has been examined 
and interpreted\cite{cov,sauls,ting2} as support for the admixture of 
an imaginary
PP component to the dominant $d_{x^2-y^2}$-wave part, leading to a broken 
time-reversal symmetry.\cite{shiba,rice}

The same developments, and the ability to make low interfacial
resistance junctions between high spin polarization
ferromagnets and superconductors,
have stimulated significant efforts to study transport in these 
structures.\cite{prinz} There have been 
various experiments in both
conventional\cite{soul,up,john} and high temperature 
superconductors\cite{vas,dong,vas2,chen} (HTSC's),
as well as re-examinations of earlier work\cite{ted,mers} which was
performed generally in the 
tunneling limit of strong interfacial barrier. 
Theoretical studies of
the effects of spin polarized transport on the current-voltage characteristics 
and the conductance in ferromagnet/superconductor (F/S)
junctions have been carried out 
 in conventional\cite{been} 
and, recently, in high-temperature superconductors.\cite{ting,zv,kash}
The feasibility of nanofabricating F/S structures 
has also
generated interest in studying the influence of ferromagnetism on mesoscopic
superconductivity.\cite{vol}

One of the important questions raised by the  possibility of making high
transmissivity F/S junctions was that of studying the influence of Andreev
reflection (AR)\cite{been,and,bru,nat} on spin polarized transport. In AR 
an electron, belonging to one of the two spin bands, 
incoming from the ferromagnetic region to the F/S 
interface will be reflected as a hole  in the opposite spin band.
The splitting of spin bands by the exchange energy in ferromagnetic 
materials implies that only a fraction of the incoming incident
majority spin electrons can be Andreev reflected.\cite{been}
This simple argument was used in  
previous studies\cite{soul,up,been} to infer that
the effect of spin polarization (exchange energy) 
was generally to reduce AR. The sensitivity of AR to
the exchange energy in a ferromagnet was employed\cite{soul,up} to determine 
the degree of polarization in various materials.

In this paper we will study the tunneling spectroscopy of F/S junctions.
We will adopt the basic
approach of Ref. \onlinecite{btk} but we will extend and
generalize it to include the
effects of spin polarization,  the presence of an unconventional PP state
(pure or mixed), and the existence of Fermi wavevector
mismatch (FWM)\cite{btk2,dutch}
stemming from the different band widths in the two junction materials.
Our aim in this paper is twofold: Firstly, to investigate and 
reveal novel features 
in the conductance spectra arising from the interplay of ferromagnetism and
unconventional superconductivity, and secondly, to show the importance of FWM, and 
how its inclusion can lead to some unexpected results, even for  F/s-wave
superconductor junctions, where we find, for example, that in some cases
Andreev reflection can be enhanced by spin polarization.

In the next section (Sec. \ref{meth}), we present the  methods we use
to obtain the amplitudes for the  various scattering processes 
that occur in the junction when spin polarized
electrons are injected from the F into the S region. We will use these
methods to calculate the conductance
of the F/S junctions. In Sec. \ref{results}, we first give results for 
a conventional
($s$-wave) superconductor in the S side,
and then illustrate  the unconventional case of the
pairing potential by considering both pure $d$- and mixed
$d+is$-wave symmetry. In 
Sec. \ref{conc},
we summarize our results and discuss future problems.
\section{Methods}
\label{meth}
As explained in the Introduction,
we investigate in this work F/S junctions by extending and generalizing
the techniques previously employed in the study of simpler
cases without spin polarization, or for conventional superconductors.
Thus, we use here the Bogoliubov-de Gennes 
(BdG) equations\cite{hu,tan,been,bru,deg} in the ballistic limit. 
We consider a geometry where the ferromagnetic material is at $x<0$, and
is described by the 
Stoner model. We take the usual approach\cite{been} of assuming a
single particle Hamiltonian
with the exchange energy being therefore of the form
$h({\bf r})=h_0 \Theta(-x)$,  where $\Theta(x)$ is a step 
function. The F/S interface is at $x=0$, where there is  
interfacial scattering modeled by a potential 
$V({\bf r})=H \delta(x)$,\cite{tan,soul,up,btk} and  $H$ is the 
variable strength of the 
potential barrier. The dimensionless parameter characterizing barrier
strength\cite{btk} is $Z_0\equiv mH/\hbar k_F$, where the
effective mass, $m$, is taken to be equal in the F and S regions.
In the superconducting region, at $x>0$, we 
assume\cite{hu,tan,been,btk,arnold} 
that there is a pair 
potential  $\Delta({\bf k}', {\bf r})=
\Delta({\bf \hat{k}}') \Theta(x)$.
This approximation for the PP becomes more accurate\cite{simple}
in the presence of FWM and allows analytic solution of the BdG equations.
We will denote quantities 
pertaining to the S region by primed letters. 

 From these considerations, the BdG equations for F/S junction, 
in the absence
spin-flip scattering, can be written as\cite{deg} 
\begin{eqnarray}
\left[\begin{array}{cc} H_0-\rho_S h & \Delta \\
\Delta^* & -(H_0+\rho_S h)\end{array} \right] 
\left[ \begin{array}{c} u_S \\
v_{\overline{S}} \end{array} \right]=\epsilon
\left[ \begin{array}{c} u_S \\
v_{\overline{S}} \end{array} \right],
\label{BdG}
\end{eqnarray} 
where $H_0$ is the single particle Hamiltonian and  $\rho_S=\pm1$ for
spin $S=\uparrow, \downarrow$. The exchange energy $h{(\bf r})$ and 
the PP $\Delta$ are as defined above. The excitation energy is denoted by
$\epsilon$, and $u_S$, $v_{\overline{S}}$ are the 
electronlike quasiparticle (ELQ)
and holelike quasiparticle (HLQ) amplitudes, respectively.
We take
$H_0\equiv -\hbar^2\nabla^2/2m+V({\bf r})-E_F^{F,S}$, where $V({\bf r})$ is
defined above. In  the F region,
we have $E_F^F\equiv E_F = \hbar^2 k_F^2/2m$, so that $E_F$ is the spin
averaged value,
$E_F=(\hbar^2 k_{F\uparrow}^2/2m+\hbar^2 k^2_{F\downarrow}/2m )/2$.
We assume that in general it differs from 
the value in the superconductor, 
$E_F^S \equiv E'_F = \hbar^2 k'^2_F/2m$. Thus,  
we take the Fermi energies to be different in the F and S regions
that is,  we allow for different 
band widths, stemming from the different carrier densities
in the two regions. Indeed, 
as the results in the next Section will show, the
Fermi wavevector mismatch (FWM) between the two regions 
has an important influence on our findings.
We will parameterize the FWM by
the value of $L_0$, $L_0 \equiv k'_F/k_F$ and describe the degree of 
spin polarization, related to the exchange energy, by the dimensionless
parameter $X\equiv h_0/E_F$. 
 
The invariance of the Hamiltonian with respect to  translations 
parallel to $x=0$ implies conservation\cite{mil} of the
(spin dependent) parallel component of the the wavevector at the junction.
As we shall show, this will be an important 
consideration in understanding the possible scattering processes.
An electron 
injected from the F side, with  spin $S=\uparrow,
\downarrow$, excitation energy $\epsilon$, and wavevector
${\bf k}^+_S$ (with magnitude
 $k^+_{S}=(2m/\hbar^2)^{1/2} [E_F +\epsilon+\rho_S h_0]^{1/2}$), 
at an angle $\theta$ from the interface
normal,  can undergo
four scattering processes \cite{tan,btk} each described by a
different amplitude. Assuming specular reflection at the interface, these can
be characterized as follows:
1) Andreev reflection,  with amplitude that we denote by $a_S$, 
as a hole with 
spin, $\overline{S}$, belonging to the spin band opposite to that of the
incident electron ($\rho_{\overline{S}}=-\rho_S$), wavevector
${\bf k}_{\overline{S}}^-$
($k^-_{\overline{S}}=(2m/\hbar^2)^{1/2} 
[E_F -\epsilon+\rho_{\overline{S}} h_0]^{1/2}$), 
and spin dependent angle of reflection $\theta_{\overline{S}}$, generally 
different  
from $\theta$.\cite{zv} As is the case with the angles 
corresponding to the other scattering processes, 
$\theta_{\overline{S}}$, as we shall see below,
is determined from  the requirement that the
parallel component of the wavevector is conserved. 
Even in the absence of exchange energy ($h_0=0$), 
one has that, 
for $\epsilon \neq 0$,
$\theta_{\overline{S}}$ (although then spin independent)
is slightly different\cite{kummel} from $\theta$. When $h_0>0$, the
typical situation is, as we 
discuss later, that
$|\theta_{\overline{\downarrow}}| <|\theta| 
< |\theta_{\overline{\uparrow}}|$.  
2) The second process is ordinary 
reflection into the F region, characterized by an amplitude
which we call $b_S$, as
an electron with variables $S$, $-{\bf k}^+_{S}$, $-\theta$.
The other two processes are:  
3) Transmission  into the S region, with amplitude $c_S$, 
as an ELQ with ${\bf k}'^+_S$, and 4) Transmission as a HLQ with
amplitude $d_S$ and wavevector $-{\bf k}'^-_S$. Here the corresponding 
wavevector magnitudes 
are $k'^{\pm}_{S}=(2m/\hbar^2)^{1/2}
[E'_F\pm(\epsilon^2-|\Delta_{S\pm}|^2)^{1/2}]^{1/2}$. We
denote by $\Delta_{S\pm}
=|\Delta_{S\pm}|\exp(i\phi_{S\pm})$,
the different  PP's  felt by the ELQ and the HLQ, respectively,
as determined by 
${\bf k}'^\pm_S$. We see, therefore, that  up to four different energy 
scales of the PP are involved for each incident angle $\theta$.
In our considerations, which pertain to the common experimental situation,
\cite{amg,venky}
$E_F, E'_F $ $\gg$ $\max(\epsilon, |\Delta_{S\pm}|)$, 
we can employ the Andreev 
approximation\cite{tan,been,and,bru} and write
$k^{\pm}_{S} \approx k_{FS}
\equiv(2m/\hbar^2)^{1/2}
[E_F +\rho_S h_0]^{1/2}$,
$k'^{\pm}_{S} \approx k'_F$. It then follows that the appropriate wavevectors
for the transmission of  ELQ's and HLQ's are at 
angles $\theta'_S$, $-\theta'_S$, 
with the interface normal, respectively.
Within this approximation the components of 
the vectors ${\bf k}^{\pm}_{S}$, 
${\bf k}'^{\pm}_{S}$
normal and parallel to the interface, can be expressed as
${\bf k}^{\pm}_{S}\equiv(k_{S}, k_{\|S})$, and
${\bf k}'^{\pm}_{S}\equiv(k'_S, k_{\|S})$,
in the F and S regions. From the conservation of $k_{\|S}$, we have then
an analogue of Snell's law
\begin{mathletters}
\begin{equation}
k_{FS}\sin\theta=k_{F\overline{S}}\sin\theta_{\overline{S}},
\label{snella}
\end{equation}
\begin{equation}
k_{FS}\sin\theta=k'_F\sin\theta'_S,
\label{snellb}
\end{equation}
\label{snell}
\end{mathletters}
which has several important implications, including the existence
of critical angles,\cite{crit}  as one encounters in well known phenomena in 
the propagation of electromagnetic waves.\cite{jackson}

Using the conservation of $k_{\|S}$, 
 the solution to Eq. (\ref{BdG}), 
$\Psi_S\equiv(u_S,v_{\overline{S}})^T$, can be expressed in a separable form,
effectively reducing the problem to a one-dimensional one. In the F region
we write
\begin{equation}
\Psi_S({\bf r})\equiv e^{i {\bf k}_{\| S}\cdot {\bf r}}\psi_S(x),
\label{Psif}
\end{equation}
where
\begin{eqnarray}
\psi_S(x)=
e^{i  k_S  x} \left[\begin{array}{c} 1 \\ 0 \end{array} \right]+
a_S e^{i k_{\overline{S}}x} \left[\begin{array}{c} 0 \\ 1 \end{array} \right] 
+b_S e^{-i k_S x} \left[\begin{array}{c} 1 \\ 0 \end{array} \right], 
\label{psif} 
\end{eqnarray}
analogously, in the $S$ region we have\cite{tan}
\begin{equation}
\Psi'_S({\bf r})\equiv e^{i {\bf k}_{\| S}\cdot {\bf r}}\psi'_S(x),
\label{Psis}
\end{equation}
\begin{eqnarray}
\psi'_S(x)=
c_S e^{i k'_S x} \left[\begin{array}{c} 
(\epsilon +\Omega_{S+}/2 \epsilon)^{\frac{1}{2}} \\ 
e^{-i\phi_+} (\epsilon -\Omega_{S+}/2 \epsilon)^{\frac{1}{2}}
 \end{array} \right] 
+d_S e^{-i k'_S x} \left[\begin{array}{c} 
e^{i\phi_-} (\epsilon -\Omega_{S-}/2 \epsilon)^{\frac{1}{2}} \\  
(\epsilon +\Omega_{S-}/2 \epsilon)^{\frac{1}{2}}
\end{array} \right], 
\label{psis} 
\end{eqnarray}
with $\Omega_{S\pm}\equiv(\epsilon^2-|\Delta_{S\pm}|^2)^{\frac{1}{2}}$,
and the appropriate boundary conditions\cite{tan,btk} at the F/S
interface are
\begin{equation}
\psi_S(0)=\psi'_S(0), \quad
\partial_x \psi_S(0)-\partial_x \psi'_S(0)=\frac{2mH}{\hbar^2}\psi'_S(0).
\label{bc}
\end{equation}

We pause next to discuss some implications of Eq. (\ref{snell}) 
for the various
scattering processes. In typical realizations of ferromagnet/HTSC structures,
the appropriate FWM corresponds to $L_0 \leq 1$.\cite{amg}
Consider first $L_0=1$, i.e. $E_F=E'_F$. If $X>0$ it follows 
that $k_{F\downarrow}<k'_F<k_{F\uparrow}$, for an $S=\downarrow$ 
incoming electron. Then, 
at any incident angle, Eq. (\ref{snell}) is satisfied so that $k_\|$ will be
conserved. In this case  $|\theta|>|\theta'_\downarrow|>|\theta_{\overline
{\downarrow}}|$, and all the corresponding wave vectors are real.
For an $S=\uparrow$ incident electron at angle 
$|\theta| > |\sin^{-1}(k'_F/k_{F\uparrow})|$,  a solution of 
Eq. (\ref{snellb}) for a real $\theta'_\uparrow$ no longer exist, one
has   
a complex $\theta'_{\uparrow}$.\cite{jackson}
The scattering problem does not have a solution with propagating
wavevectors in the S region:
there is total reflection. 
The  wavevectors for
ELQ and HLQ have purely imaginary components along the $x$-axis, while
their components parallel to the interface  are real.
This corresponds to  a surface (evanescent) wave, propagating 
along the interface and exponentially damped away from 
it.\cite{jackson} 
An analogous, but physically more interesting, situation occurs for AR in  
the particular case where
$|\theta|$ is smaller than the angle of total reflection  and satisfies
$|\theta|>|\sin^{-1}(k_{F\downarrow}/k_{F\uparrow})|$. This regime 
corresponds to 
$k_{\|\overline{\uparrow}} > k_{F\downarrow}$. In this case it is
Eq. (\ref{snella}) that has no solution for real angles. This means
that Andreev reflection as a {\it propagating} wave is impossible. 
 From the condition, which follows from the Andreev approximation, 
$k_{\overline{\uparrow}}^2+k_{\|\overline{\uparrow}}^2
\equiv k_{F\downarrow}^2$,  we see that the 
component $k_{\overline{\uparrow}}$ along the $x$ axis must
be purely imaginary,\cite{kash} while 
$k_{\|\overline{\uparrow}}$ is still real. 
With these considerations we then find 
\begin{equation}
k_{\overline{\uparrow}}=-i 
(k_{F\uparrow}^2 \sin^2\theta-k_{F\downarrow}^2)^{1/2},
\label{kim}
\end{equation}
where we have expressed 
$k_{\overline{\uparrow}}$ in terms of quantities which are always real
and which pertain to the F region only. As with total reflection, 
there is 
propagation only along the interface and an exponential decay away from it.
This case differs from that
of total reflection in that, since the evanescence affects only the
Andreev reflected component, there may still be
transmission across the junction.

The above considerations apply {\it a fortiori} in the
presence of FWM. For example, if we now  consider $L_0<1$, we can see
by inspection of Eq. (\ref{snell}),
that there can also be total reflection for an
$S=\downarrow$ incident electron, when $k_{F\downarrow}>k'_F$. This
condition  would imply the  absence of imaginary 
$k_{\overline{\uparrow}}$ for any incident angle and any exchange energy.

Returning now to the basic equations, we see that by solving for   
$\psi_S(x)$, $\psi'_S(x)$ in Eq. (\ref{psif}), (\ref{psis}) with
the boundary conditions given by Eq. (\ref{bc}), we can obtain the
amplitudes $a_S$, $b_S$, $c_S$  and $d_S$, $S=\uparrow,\downarrow$.
For each spin, there is a sum rule, related to the conservation of probability, 
for the squares  of the absolute 
values of the amplitudes. 
We can thus, in a way similar  to what was done in
Ref. \onlinecite{btk}, express the various quantities in terms of the 
amplitudes $a_S$ and $b_S$ only. These amplitudes are given by
\begin{equation}
a_S=\frac{4 t_S L_S \Gamma_+ e^{-i\phi_{S+}}}
{U_{SS+}U_{\overline{S} S-}-
V_{SS-} V_{\overline{S} S+}
\Gamma_+\Gamma_-e^{i(\phi_{S-}-\phi_{S+})}},
\label{as}
\end{equation}
\begin{equation}
b_S=   \frac
{V_{S S+}U_{\overline{S} S-}-
U_{S S-} V_{\overline{S} S+}
\Gamma_+\Gamma_-e^{i(\phi_{S-}-\phi_{S+})}}
{U_{SS+} U_{\overline{S} S-}-
V_{SS-} V_{\overline{S} S+}
\Gamma_+\Gamma_-e^{i(\phi_{S-}-\phi_{S+})}},
\label{bs}
\end{equation}
where we have introduced the notation
$\Gamma_\pm\equiv
(\epsilon-\Omega_{S\pm})/|\Delta_{S\pm}|$,
$L_S\equiv L_0 \cos\theta'_S/\cos \theta$, 
describing FWM,
$t_S\equiv k_S/k_{Fx}=(1+\rho_S X)^{1/2}$,
$t_{\overline{S}}\equiv k_{\overline{S}}/k_{Fx}=
(1-\rho_S X)^{1/2} \cos \theta_{\overline{S}} /\cos \theta$, for 
$k_{\overline{S}}$ real, ($-i[(1+X)\sin^2\theta-(1-X)]^{1/2}/\cos\theta$, for 
$k_{\overline{\uparrow}}$ imaginary, see Eq. (\ref{kim})).
The other abbreviations are defined as:
$U_{\overline{S} S\pm}\equiv t_{\overline{S}}+w_{S\pm}$,
$V_{S S\pm}\equiv t_{S}-w_{S\pm}$,
 $w_{S\pm}\equiv L_S\pm 2iZ$, $Z\equiv Z_0/\cos\theta$, where
$Z_0\equiv m H/\hbar k_F$  is the interfacial barrier parameter,
as defined above.
The limits  $Z_0 \rightarrow 0$ and
 $Z_0 \rightarrow \infty$ correspond to the extreme cases of a metallic
point contact and the tunnel junction limit, respectively.

Given the above amplitudes, the results for the dimensionless differential
conductance\cite{btk} can  be written down in the standard way
by computing, as a function of the excitation energy arising
from the application of a bias voltage, the ratio of the induced flux
densities across the junction  
to the corresponding incident
flux density. One straightforwardly generalizes the
methods used in previous work\cite{tan,btk,aver} 
to include now the
effects of unconventional
superconductivity, FWM,  and net spin polarization,
to obtain,
\begin{equation}
G\equiv G_{\uparrow}+G_{\downarrow}=\sum_{S=\uparrow,\downarrow} P_S
(1+\frac{k_{\overline{S}}}{k_S}|a_S|^2-|b_S|^2),
\label{gs}
\end{equation}
where we introduce
the probability $P_S$ of an incident electron having spin $S$, 
related to the exchange energy as $P_S=(1+\rho_SX)/2$.\cite{been}
In deriving Eq. (\ref{gs}), care has to be taken to 
properly include the flux factors, which
are, at $X>0$,  different for the incident and the Andreev reflected particle.
The ratio of wavevectors in the second term on
the right side of Eq. (\ref{gs}) results from  the incident electron and the AR
hole belonging to different spin bands.
The quantity $k_{\overline{S}}$ in that term 
is  real, the case
of imaginary $k_{\overline{S}}$ can only contribute to $G_\uparrow$ indirectly,
by modifying $|b_\uparrow|$.
 It  can be shown\cite{zv}
from the conservation of probability current\cite{btk} that
such a contribution vanishes  
for the subgap conductance ($\epsilon<|\Delta_{\uparrow \pm}|$).\cite{total} It
is, furthermore, possible to express 
the subgap conductance in terms of the AR amplitude only.\cite{zv}
At $X=0$ we recover the results of  Ref.  \onlinecite{tan}.
The suppression  of  the conductance due to  ordinary reflection 
at $X\neq 0$ has the same
form as for the unpolarized case since the magnitude of the normal component of
the wavevectors before and after ordinary reflection remains the same.

We focus in this work (see results in the next Section) on the conductance
spectrum of the charge current as given by Eq. (\ref{gs}),
but the amplitudes $a_S$, $b_S$, given by Eq. (\ref{as}), (\ref{bs})
can be used to calculate many other quantities of interest, such as 
current-voltage characteristics, the spin current,
and the spin conductance.\cite{kash} We consider also here
angularly averaged quantities and notice 
that  Eq. (\ref{gs}) implies that the conductance 
vanishes for $|\theta|$ 
greater than the angle of total reflection (we recall that this angle
is spin dependent). We define the angularly averaged
(AA) conductance, $\langle G_S\rangle$, as
\begin{equation}
\langle G_S \rangle=\int_{\Omega_S} d\theta \cos \theta G_S(\theta)
/\int_{\Omega_S} d\theta \cos\theta,
\label{ga}
\end{equation}
where $\Omega_S$ is limited by the angle of total reflection
or by experimental setup. This form correctly reduces to 
that used in the previously
investigated spin unpolarized situation.\cite{tan} One may choose
a different weight 
function in 
performing such angular averages, depending
on the specific experimental geometry 
and the strengths of the interfacial 
scattering.\cite{wei,aver,sauls2}
However, all expressions for angularly averaged results, obtained
from  different averaging methods, 
would still have a factor of 
$(1+\frac{k_{\overline{S}}}{k_S}|a_S|^2-|b_S|^2)$ in the kernel of 
integration, and would merely require numerical integration of the
amplitudes we have already given here.

\section{Results}
\label{results}
\subsection{Conventional pair potentials}
\label{cpp}
We present our results in terms of the dimensionless
differential conductance, plotted  as a function of the
dimensionless energy $E\equiv \epsilon/\Delta_0$. We
concentrate on the region $E\lesssim 1$ since for larger bias 
various extrinsic effects, such as heating, tend to dominate the behavior
of the measured conductance.\cite{vas3} While our findings,
and the analytic results from Section \ref{meth}, are valid for any value 
of the interfacial scattering, we focus on smaller values of $Z_0$,
$Z_0\leq 1$, where the novel effects of ferromagnetism on Andreev reflection,
and consequently on the conductance, are  more pronounced than in the tunneling
limit, $Z_0 \gg 1$. This regime on which we focus is also that
which is believed to
correspond to several ongoing experiments of F/S structures, where the 
samples typically have  small interface resistance.\cite{vas2,chen}
To present numerical results, we choose $E'_F/\Delta_0=12.5$, consistent with 
optimally doped
$YBa_2Cu_3O_{7-\delta}$.\cite{kresin,hars} We will include
FWM, as parametrized by the quantity $L_0$ introduced above, 
$E_F=E'_F/L_0^2$.

We first give some results for an $s$-wave PP, with a constant energy gap.
This will serve to illustrate the influence of FWM coupled with 
that of $Z_0$
within a simpler and more familiar context.
In this case,
for any incident angle, $\theta$, of an injected electron the ELQ and HLQ 
feel the same PP with $\Delta_{S\pm}=\Delta_0$, and $\phi_{S\pm}\equiv 0$.
Therefore, the results that we give here for the $s$- wave case and 
normal incidence ($\theta=0$),
also correspond to the case of  a PP of the  $d_{x^2-y^2}$ form, with
the  angle $\alpha \in (-\pi/2,\pi/2)$,  between the crystallographic
$a$-axis and the interface normal, set to $\alpha=0$.  This would
represent an F/S interface along the (100) plane.

\begin{figure}[htbp]
\vspace*{-2.0cm} \hspace*{4cm}
\epsfxsize = 3.4 in \epsfbox{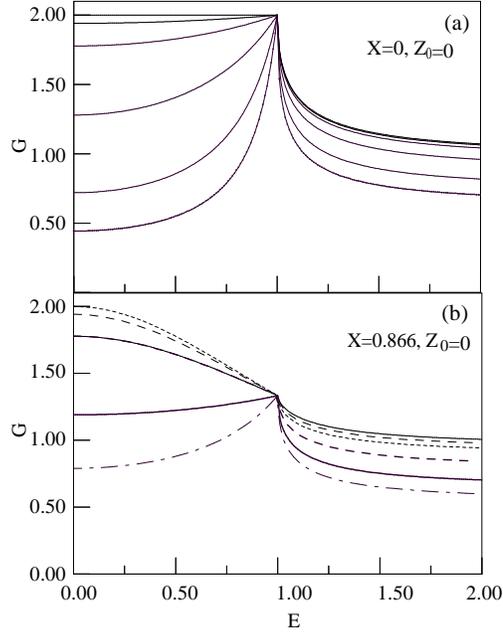}
\vspace*{-0.75cm}
\caption{$G(E)$ (Eq.(\protect{\ref{gs}})) versus
$E\equiv \epsilon/\Delta_0$. Results are for $\theta=0$ (normal incidence).
The  curves are  for $Z_0=0$ (no barrier):
in panel (a) at  exchange energy $X\equiv h_0/E_F=0$ (no spin polarization) 
they are (from top to bottom at any $E$) 
for the FWM values of $L_0^2=E'_F/E_F=1,1/\sqrt{2},1/2,1/4,1/9,1/16$.
In panel (b) they are for $X=0.866$. Since the curves now cross at $E=1$
they are drawn in different ways for clarity.
For $E>1$ they are in the same order as in panel (a) and
for the same values of $L_0$,
while for $E<1$ they correspond, from top to bottom, 
to $L_0^2=1/2,1/\sqrt{2},1,1/9,1/16$. The $L_0^2=1/4$ curve
overlaps with that for $L_0^2=1$ in this range.}\label{l1}
\end{figure}                   

In Fig. \ref{l1} we show results for $G(E)$, given by Eq. (\ref{gs}), at
$\theta=0$,  and $Z_0=0$ 
(this limit of no  interfacial barrier was also considered 
in Ref.\onlinecite{been}). We plot results for various values of  the FWM 
parameter $L_0$. Panel (a) corresponds to no polarization ($X=0$)
and panel (b) to high polarization $X=\sqrt{3}/2\approx 0.866$. 
For normal incidence, 
we have 
$t_S=(1+\rho_S X)^{1/2}$,
$t_{\overline{S}}= (1-\rho_S X)^{1/2}$ (as defined below Eq. (\ref{bs})), and
the subgap conductance can be expressed as
\begin{equation}
G=\frac{32 L^2_0 (1-X^2)^{1/2}}
{|t_\uparrow t_\downarrow+ (t_\uparrow+t_\downarrow)L_0+L^2_0
-(t_\uparrow t_\downarrow- (t_\uparrow+t_\downarrow) L_0+L_0^2)
\Gamma_+ \Gamma_-|^2}. 
\label{sub}
\end{equation}
Panel (a) displays results in the absence of exchange energy.
With  increasing
FWM (i.e. decreasing $L_0$), the amplitude at zero bias voltage (AZB) decreases
monotonically. This effect was explained\cite{btk2} in previous work
as resulting from the increase  in a single parameter 
$Z_{eff}$, which combined $Z_0$ with the effects of FWM. 
Our curves with  FWM ($L_0 < 1$) reduce in the appropriate limits to those 
previously found\cite{btk2} with $L_0=1$
and $Z_0 \rightarrow Z_{eff}$, $Z_{eff}>Z_0$. We will see below that this
is not the case at $X\neq 0$.
In panel (b) we give results for high $X$
while keeping the
other parameters at the same values as in panel (a). We notice that the presence of
exchange energy gives rise to non-monotonic behavior in the  AZB. At  low
bias, the conductance can be enhanced with increasing FWM (compare,
for example, the $L_0=1$ and $L_0=1/\sqrt{2}$ results),
and form a zero bias conductance peak (ZBCP.) This behavior is qualitatively 
different from that found in the 
unpolarized case and the effect of FWM can no longer be reproduced
by simply increasing the interface scattering parameter.
Thus the often implied\cite{soul,up} expectation that the effects of $Z_0$ and $L_0$ 
could also be 
subsumed in a single parameter in the spin polarized case is not fulfilled. 
In this panel we have an example of coinciding subgap conductances 
for $L_0=1$ and $L_0=1/2$.
The condition for this coincidence to take place at
fixed X can be simply obtained from Eq. (\ref{sub}) as
\begin{equation}
t_\uparrow t_\downarrow/L^2_0=L'^2_0 \quad \Rightarrow \quad 
(1-X^2)=L'^2_0 \: (L_0\equiv1),
\label{coincide}
\end{equation}
where $L_0$, $L'_0$ correspond to two different values of FWM for 
which the subgap conductances will coincide.

We next look, in the same situation as in the previous figure,
at the effects of the presence of an
interfacial barrier. 

\begin{figure}[htbp]
\vspace*{-2.0cm} \hspace*{4cm}
\epsfxsize = 3.4 in \epsfbox{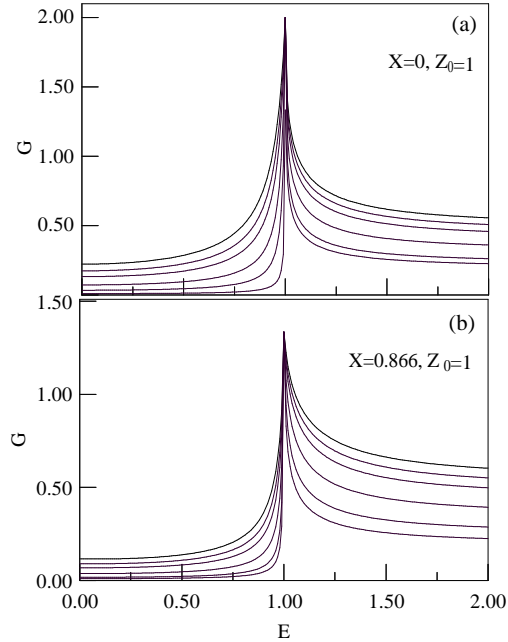}
\vspace*{-0.75cm}
\caption{$G(E)$ for $\theta=0$ and  
interfacial barrier strength $Z_0=1$. All the other  parameters 
are taken as in the previous figure. In both panels, 
curves from top to bottom correspond  to decreasing values of $L_0$.}
\label{l2}
\end{figure}

In Fig. \ref{l2}, we choose $Z_0=1$,
while keeping all the other parameters the same as in
the corresponding panel of the previous figure. In panel (a) 
we show results in the absence of spin 
polarization. A finite bias conductance peak
(FBCP) appears at the gap edge. It becomes increasingly narrow with
greater FWM (smaller $L_0$). 
Its amplitude
is $2$, independent of $L_0$. In panel (b), at $X=0.866$, the conductance
curves display similar behavior, but  with a reduced FBCP at the gap
edge. From Eqs. (\ref{as}), (\ref{bs}), and (\ref{gs}), the amplitude
of the FBCP in this case is
\begin{equation}
G(E=1)=\frac{4(1-X^2)^{1/2}}{1+(1-X^2)^{1/2}}.
\label{edge}
\end{equation}
An interesting feature of this result is that it depends only
on the exchange energy
(spin polarization) and not on the FWM parameter
or the barrier strength. It can be shown that this property holds
for all angles of incidence. This is in contrast 
with the  value of the zero bias conductance which depends,  both for normal
incidence and for other angles,  on the value of the FWM. This dependence 
could introduce  difficulties in the accurate determination of spin 
polarization from the AZB.\cite{soul,up} The gap edge
value is less susceptible to these problems.

\begin{figure}[htbp]
\vspace*{-3.75cm} \hspace*{4cm}
\epsfxsize = 3.4 in \epsfbox{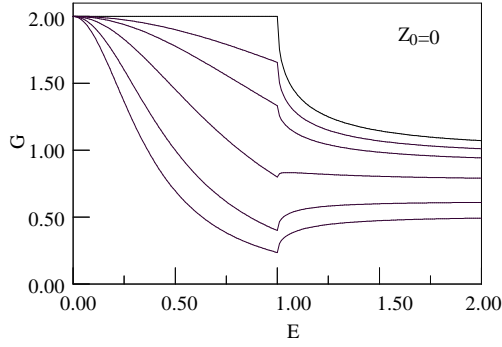}
\vspace*{-3.0cm}
\caption{Evolution of the zero bias conductance, $G(E)$  for $\theta=0$. 
Results are given  at $Z_0=1$ for $X$ determined from Eq. (\protect\ref{azb}) 
 and values of $L_0$ as in Fig. \protect{\ref{l1}}. From top to
bottom the curves correspond to
values of $(L_0^2,X)$ given by $(1,0)$, $(1/\sqrt{2},1/\sqrt{2})$,
$(1/2,0.866)$, $(1/4,0.968)$, $(1/9,0.994)$, and $(1/16,0.998)$.}
\label{l3}
\end{figure}

The presence of spin polarized carriers, due to nonvanishing
exchange energy, is usually held\cite{up,been,ting} to result in the suppression
 of Andreev reflection and thus  in a  reduction of the subgap
conductance. A simple explanation,\cite{been} which neglects the effects of FWM,
predicts that the AZB should monotonically decrease with increasing
$X$, because of the reduction of Andreev reflection, when
only a fraction  of injected electrons from the majority spin band
can be reflected as holes belonging to the minority spin band.
This follows from the reduction of the density of states in the 
minority spin band with increasing $X$, and eventually causes 
the subgap conductance to vanish for a half-metallic ferromagnet when 
$X\rightarrow 1$. 

\begin{figure}[htbp]
\vspace*{-2.0cm} \hspace*{4cm}
\epsfxsize = 3.4 in \epsfbox{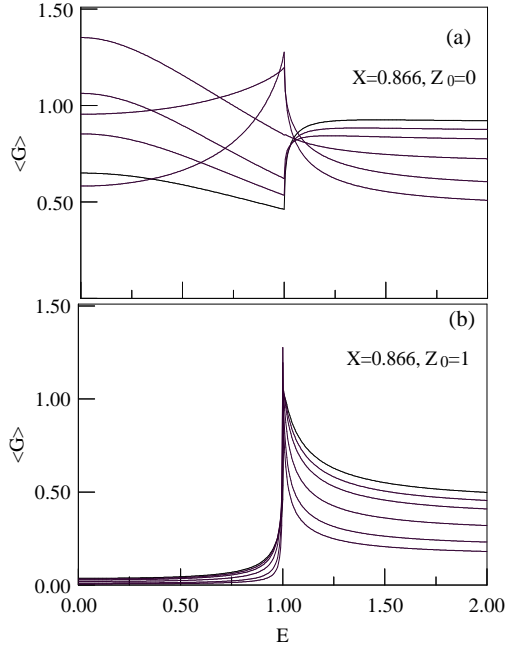}
\vspace*{-0.75cm}
\caption{$\langle G(E)\rangle$, the $\theta$ averaged
conductance, for  an $s$-wave PP and the same values  
of  $X$, $L_0$ as in panels (b) of Figs. \protect{\ref{l1}},
\protect{\ref{l2}}, respectively. 
In both panels curves from top to bottom, at
$E=2$, correspond to decreasing $L_0$.}
\label{l4}
\end{figure}

We now proceed to examine whether these findings are
modified when FWM is taken into account.
In Fig. \ref{l3}, which shows results at $Z_0=0$ and normal incidence,
we consider the evolution of the conductance
curves for different values of $X$ and $L_0$ chosen to yield maximum AZB, 
$(G(E=0)=2$), starting from the step-like feature at $L_0=1$ and $X=0$
(see Fig. (\ref{l1})).
The condition for maximum AZB at fixed FWM and polarization 
can be derived\cite{zv} from Eq. (\ref{sub}) and is
\begin{equation}
k_{\uparrow} k_{\downarrow}=k'^2_F
\quad \Rightarrow \quad(1-X^2)^{1/2}=L_0^2.
\label{azb}
\end{equation}
We have used this equation to determine the optimal value of $X$ for each 
value of $L_0$ used
in Figs. \ref{l1}, \ref{l2}. The resulting curves are plotted 
in Fig. \ref{l3}.
This figure reveals several interesting features. With the increase
of FWM  and the correspondingly larger 
optimal spin polarization (according
to the value of $X$ found from Eq. (\ref{azb})), a ZBCP forms.
This is a novel effect in which the peak arises from a mechanism completely 
different
from the one usually put forward, where the ZBCP
is  attributed to  the 
presence of unconventional superconductivity. In that case, the ZBCP
is produced by the sign change of the PP and the concomitant
formation of Andreev bound 
states.\cite{hu,tan,alf}
Furthermore, if we compare these curves with those in panel (b) of Fig. \ref{l1},
we see that the subgap conductance can increase with increasing 
spin polarization at fixed $L_0$. This implies that Andreev
reflection can be enhanced by spin polarization. 

We now turn to angular averages (AA).
In Fig. \ref{l4} we show angularly averaged results, obtained
from the  expression  for $\langle G \rangle$,
Eq. (\ref{ga}). The averaged results are no longer equivalent
(as in the previous figures with normal incidence) to the case of 
a $d_{x^2-y^2}$ PP with an  F/S interface along the (100) plane: the angular
dependence of the PP would then modify the results.
Each of the two panels shown includes results for the same set of parameter
values 
used in panels (b) of Figs. \ref{l1}, and \ref{l2}, respectively.
In panel (a) of the current figure we show how the novel features
previously discussed are largely
preserved after angular averaging. There is still formation
of a ZBCP with increased FWM and the AZB retains its non-monotonic
behavior with $L_0$, as in the case of fixed normal incidence.
The angularly averaged results in panel (b), at $Z_0=1$,
display behavior similar to that found in the $\theta=0$ case, 
with the conductance
peak at $E=1$ becoming sharper at increasing FWM. 

\subsection{Unconventional pair potentials}
\label{upp}
We next consider an angularly dependent PP, specifically that
for a $d_{x^2-y^2}$ pairing state. With this PP
we have different, spin dependent,  PP's for ELQ's and HLQ's. These
are given respectively by
$\Delta_{S\pm}=\Delta_0\cos(2 \theta'_{S\pm})$, where $\theta'_{S\pm}$
can be expressed as
$\theta'_{S\pm}=\theta'_{S}\mp\alpha$ (we recall  that $\alpha$ is
the angle between the interface normal and the crystallographic $a$-axis,
and $\theta'_S$ is related  to $\theta$ through Eq. (\ref{snell})).

\begin{figure}[htbp]
\vspace*{-2.0cm} \hspace*{4cm}
\epsfxsize = 3.4 in \epsfbox{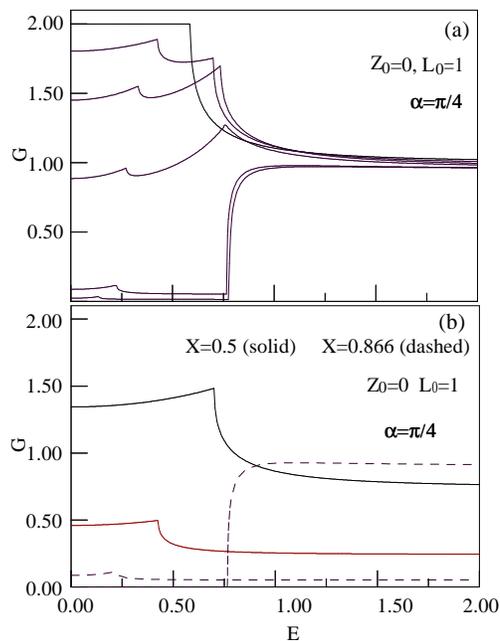}
\vspace*{-0.75cm}
\caption{$G(E)$
 for $\theta=\pi/10$, $\alpha=\pi/4$, $Z_0=0$, and $L_0=1$.
In (a) the curves are for
$X=0,0.5,0.7,0.8, 0.866,0.95$,
(top to bottom at $E=0$).
In (b), we plot the 
spin resolved conductance, for two values of $X$.  
The upper curve at $E>1$ corresponds to
$G_\uparrow$, and  and lower curve to $G_\downarrow$.}
\label{l5}
\end{figure}

In Fig. \ref{l5} we give some of our results 
for $d$-wave pairing and $\alpha=\pi/4$ (interface in the (110) plane),
in the absence of both interfacial barrier and FWM and   
 at a fixed $\theta=\pi/10$, for various values of $X$.
Panel (a) shows curves  for  the total conductance as it
evolves from  a step-like
feature at $X=0$ to a zero bias conductance dip (ZBCD) for large 
spin polarization. The width of the plateau at $X=0$ is determined by a
single energy scale set by the equal magnitudes of the PP's for ELQ and HLQ
in that case, as given by
$\Delta_{S+}=\Delta_{S-}<\Delta_0$, $S=\uparrow,\downarrow$. As the 
exchange energy is increased, $k_{F\uparrow}$ and $k_{F\downarrow}$
are no longer equal. As one can see from Eq. (\ref{snellb}), it follows that 
$\theta'_\uparrow \neq \theta'_\downarrow$ and thus 
$\Delta_{\uparrow\pm} \neq \Delta_{\downarrow\pm}$. These two different energy
scales are responsible for the position of various features, such as the
several finite bias conductance peaks (FBCP's) that are seen.

\begin{figure}[htbp]
\vspace*{-2.0cm} \hspace*{4cm}
\epsfxsize = 3.4 in \epsfbox{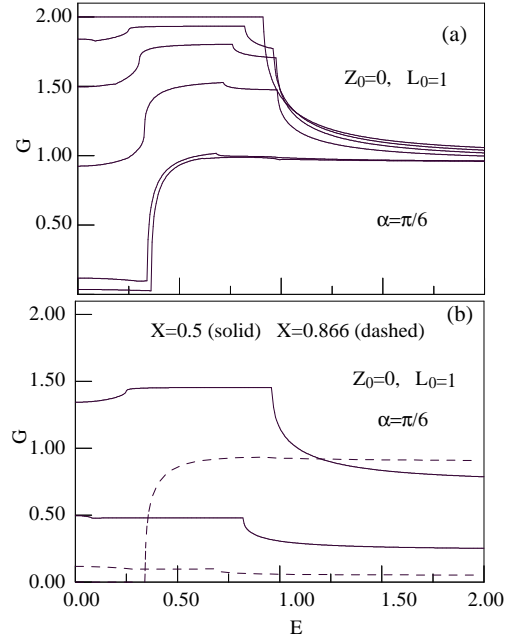}
\vspace*{-0.75cm}
\caption{$G(E)$
 for $\theta=\pi/10$, $\alpha=\pi/6$, $Z_0=0$, and $L_0=1$.
In both panels ordering and values of $X$ for each curve are as in
Fig. \protect{\ref{l5}}.}
\label{l6}
\end{figure}

In panel (b) we show the spin decomposition $G=G_\uparrow+G_\downarrow$,
which better reveals these scales, at two different exchange energies.
At $X=0.5$, the shapes of  $G_\uparrow$, $G_\downarrow$ are only slightly 
modified
from those in the unpolarized case. 

\begin{figure}[htbp]
\vspace*{-2.0cm} \hspace*{4cm}
\epsfxsize = 3.4 in \epsfbox{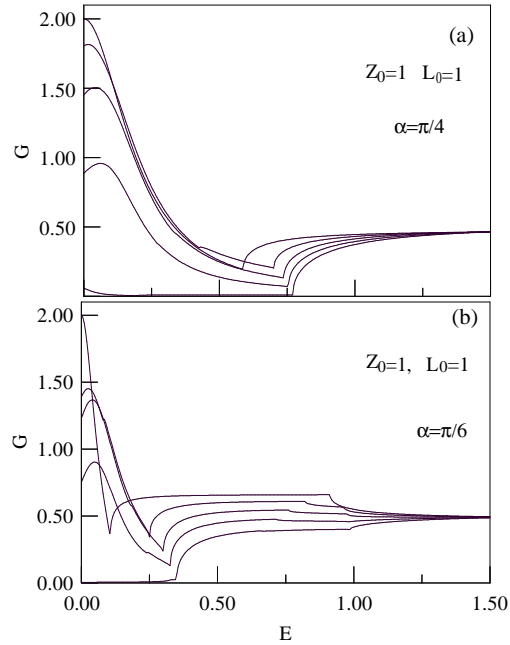}
\vspace*{-0.75cm}
\caption{$G(E)$
 for $\theta=\pi/10$, $Z_0=1$, and $L_0=1$. In both panels
curves (top to bottom at $E=0$) correspond to $X=0,0.5,0.7,0.8,0.9$.
In (a) $\alpha=\pi/4$, and in (b) $\alpha=\pi/6$.}
\label{l7}
\end{figure}

At larger exchange energy, $X=0.866$, 
the situation is 
very different, as shown in the figure. 
We also see, in
panel (b), that as 
stated in the previous section, the evanescent wave associated
with the imaginary $k_{\overline{\uparrow}}$
does not contribute to the subgap conductance $G_\uparrow$.

In general, for an arbitrary orientation of the
F/S interface, $\alpha\neq 0,\pi/4$, at a fixed $\theta$, all the four
spin dependent PP's for ELQ and HLQ will have different magnitudes.
There are, therefore, specific features at four different energy scales.
It is only for the particular and atypical (but often
chosen in theoretical work) case of $\alpha=\pi/4$
that these four scales reduce to two.

In Fig. \ref{l6} we show the general behavior by choosing $\alpha=\pi/6$,
while retaining the values of all the other parameters from the previous
figure. One can easily calculate, for example, that at $X=0.5$ the 
normalized values 
of the PP are, in units of the gap maximum, $\Delta_0$,
$|\Delta_{\uparrow+}|=0.963$, $|\Delta_{\uparrow+}|=0.250$,
$|\Delta_{\downarrow+}|=0.822$, $|\Delta_{\downarrow-}|=0.083$.
These numbers can also be approximately inferred from the spin resolved
results given by the solid lines in panel (b).

\begin{figure}[htbp]
\vspace*{-2.0cm} \hspace*{4cm}
\epsfxsize = 3.4 in \epsfbox{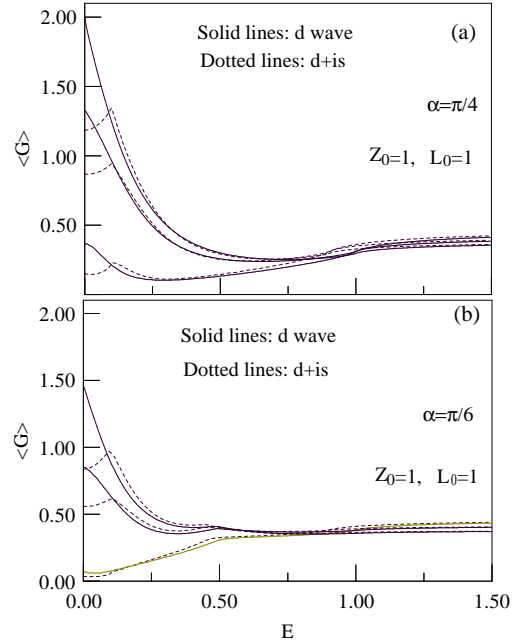}
\vspace*{-0.75cm}
\caption{$\langle G(E) \rangle$, at $Z_0=1$, and $L_0=1$ 
for $d_{x^2-y^2}$, and
 $d_{x^2-y^2}+is$ pair potentials. The latter is of the form
$\Delta_{S\pm}=\Delta_0\cos(2 \theta'_{S\pm})+i0.1\Delta_0$.
In panel (a) $\alpha=\pi/4$ and in (b) $\alpha=\pi/6$. From top to 
bottom (at $E=0$), the  curves correspond to $X=0,0.5,0.9$, 
in both panels and for each pair potential.} 
\label{l8}
\end{figure}

We next turn to the case where there is a nonvanishing potential barrier,
choosing for illustration the value $Z_0=1$. In the absence of spin 
polarization, the 
formation of a ZBCP at finite barrier strength
has been extensively investigated\cite{hu,tan,xu} 
and explained by  Andreev bound states 
in the context of $d$-wave superconductivity. We will consider here also
the effects of $X$, not included in previous work. 
In Fig. \ref{l7} we show results for various values of $X$ at $\alpha=\pi/4$.
(in panel (a)) and $\alpha=\pi/6$ (panel (b)). One can see that 
for intermediate values of $X$
the conductance maximum is at finite bias.  Comparing the two panels, 
one sees that the AZB 
at a fixed $X\neq 0$ is larger for $\alpha=\pi/4$, in agreement with the
results obtained for the
unpolarized case where, at zero bias, the spectral weight is maximal\cite{hu}
for a (110) interface.
For a different choice of incident angle $\theta$ there will be, 
if the values of all other parameters are held fixed, a change in the effective
barrier strength for various scattering processes. 
We recall 
(see below Eq. (\ref{bs})), that $Z=Z_0/\cos\theta$, and with an
increase in $|\theta|$ typically there will be,
as in the unpolarized case,\cite{bag}  a decrease in the amplitude for Andreev 
reflection and an increased amplitude for
ordinary reflection.

Results such as those discussed above can be obtained as a function of angle,
and the angular average can then be computed from Eq. (\ref{ga}). We will
combine showing some of these angularly averaged results with a brief study of 
another point:
it is straightforward to use the formalism discussed here to examine more
complicated superconducting order parameters. A question that has given
rise to a considerable amount of discussion is that of
whether the superconducting order parameter in high $T_c$ materials
is pure $d$-wave or contains a mixture of $s$ wave as well, with an
imaginary component, so that there would not be, strictly speaking, 
gap nodes, but only very deep minima. With this in mind,
the effect of a possible ``imaginary'' PP admixture (for example in a
$d+is$ form) on Andreev bound states has also been recently studied. 
\cite{cov,sauls,ting2} We consider this question here, including the
effects of polarization.
In  Fig. \ref{l8}, we illustrate the difference in the 
angularly averaged conductance values
obtained for a pure
$d_{x^2-y^2}$ PP and for a mixed  $d_{x^2-y^2}+is$ case. We choose the
particular 
form $\Delta_{S\pm}=0.9\Delta_0\cos(2 \theta'_{S\pm})+i 0.1\Delta_0$.
The phase of the PP, $\phi_{S\pm}$, is no longer equal to $\pi$ or $0$ as in
the pure $d$-wave case. We give AA results for several values 
of $X$, both for the pure $d$ and the mixed $d+is$ cases. 

\begin{figure}[htbp]
\vspace*{-2.0cm} \hspace*{4cm}
\epsfxsize = 3.4 in \epsfbox{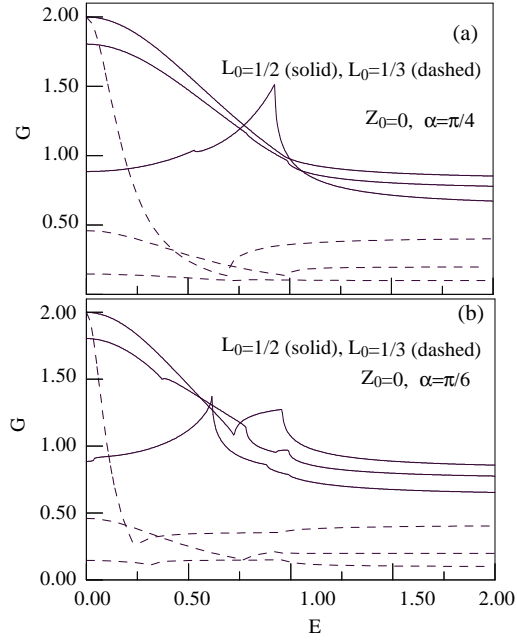}
\vspace*{-0.75cm}
\caption{$G(E)$
 for $\theta=\pi/10$, $Z_0=0$, and $L_0=1/2,L_0=1/3$. 
In panel (a) $\alpha=\pi/4$ and in (b) $\alpha=\pi/6$.
 From top to bottom, at $E=0$,  curves correspond to $X=0,0.5,0.8$, 
in both panels and for each pairing potential.} 
\label{l9}
\end{figure}

The former
represents the angular average of results similar to those 
previously displayed. 

\begin{figure}[htbp]
\vspace*{-2.0cm} \hspace*{4cm}
\epsfxsize = 3.4 in \epsfbox{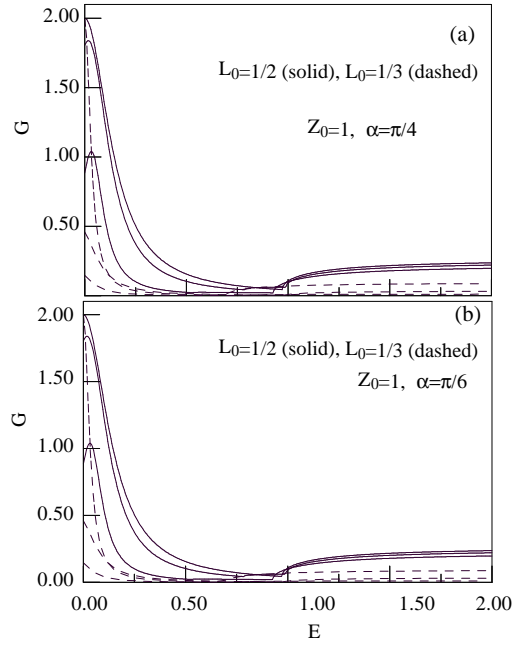}
\vspace*{-0.75cm}
\caption{Conductance curves for $\theta=\pi/10$ at $Z_0=1$ with the same
parameters and ordering as in Fig. \protect{\ref{l9}}.}
\label{l10}
\end{figure}

As in the unpolarized case,\cite{sauls,ting} the $is$ 
admixture
in the PP is responsible for a FBCP, approximately at
$E=0.1$. The conductance maximum is reduced with increased $X$ and 
with departure from  a (110) oriented interface. 
Replacing the
 $d_{x^2-y^2}+is$ PP by a ``real'' admixture   $d_{x^2-y^2}+s$ 
(taking again $0.1\Delta_0$ for the $s$-wave part)  gives 
results almost indistinguishable from the  pure $d$-wave for any
value of spin polarization. 

To show the effects of FWM on conductance for a pure $d$-wave PP we take
$L_0=1/2,1/3$ and give results at the fixed angle, $\theta=\pi/10$,
previously considered.  In Fig. \ref{l9} we show  curves at $Z_0=0$
and $\alpha=\pi/4$ (panel (a)), and for  $\alpha=\pi/6$ in panel (b).
It is useful to compare this figure to panel (a) in Figs. \ref{l5}, and
\ref{l6}, corresponding to no FWM for $\alpha=\pi/4$ and $\pi/6$, respectively.
In the absence of spin polarization the effect of FWM resembles the influence
of a nonvanishing barrier strength, $Z_0$, and leads to the formation
of a ZBCP,
which becomes increasingly narrow for smaller $L_0$. The effect of moderate
spin polarization ($X \lesssim 0.5$, for comparison with the above 
mentioned figures) on the AZB is 
rather small for $L_0=1, 1/2$ but it is significantly larger at $L_0=1/3$.
In the next figure, Fig. \ref{l10}, we use $Z_0=1$ and the same parameters as in 
the previous
figure, so that the influence of barrier strength can
be gauged.  One sees that in the presence of spin polarization the position
of the conductance maximum depends on FWM. With increasing mismatch, 
the FBCP
evolves into a ZBCP. By comparing the curves
corresponding to $L_0=1$ in Fig. \ref{l7} with those
for smaller $L_0$ in Fig. \ref{l10} , it is 
interesting to  notice that an effect  similar to that
discussed previously for $s$-wave PP without an interfacial barrier 
and at normal incidence is also manifested in other regimes, in that 
the conductance maximum can actually be enhanced,
in the spin polarized case (at fixed $X$), by the FWM.
\section{Conclusions}
\label{conc}
In this paper we have studied the conductance spectra of 
ferromagnetic/superconductor structures. The expressions for 
Andreev reflection and ordinary reflection amplitudes which we have given, allow one to
simply obtain other quantities of interest such as current-voltage 
characteristics or conductance spectra for spin current.\cite{kash}
We have developed the appropriate extensions of
the standard approach and approximations used in  
the absence of spin polarization. This has enabled us to present analytic 
results. Within these approximations, and with the inclusion of FWM,  we have 
shown a number of important
qualitative differences from the unpolarized case or from that where
spin polarization is included 
in the absence of FWM. 

Our considerations may also be important in the interpretation of recent
experiments\cite{soul,up} attempting to use tunneling to
measure the degree of spin polarization in the ferromagnetic
side of the junction, since the experimental determination
of spin polarization in a ferromagnet is a very difficult
and important experimental question in its own right.
As we have shown, the ZBCP is sensitive to both spin polarization and FWM,
while the gap edge amplitude depends only on $X$.
It is then not possible to straightforwardly determine the spin polarization 
by using the results for the amplitude of the zero bias conductance unless the 
appropriate FWM of the F/S structure is
known and properly taken into account. Furthermore, FWM can not, unlike in the 
unpolarized case,\cite{btk2}
be simply described by a rescaled 
value of the interfacial barrier strength.

The procedures used here have the advantages of simplicity and of allowing
for analytic solutions. These advantages have enabled us to investigate
widely the relevant parameter space. 
We have left  for future work considerations that would have diminished
these advantages. Among these
are the question of the  self consistent treatment of the PP, inclusion
of spin-flip scattering or of a 
more realistic band structure, and non-equilibrium transport. However,
we believe that  the the methods we have employed 
are sufficient to elucidate the hitherto unappreciated subtleties and 
the richness and variety of the phenomena associated with
spin polarized tunneling spectroscopy.

We hope that our work,
as reported here and in Ref.{\onlinecite{zv}, will prompt additional experiments 
and theoretical work. In particular,
an important clue about spin polarized transport would be provided by
measurements of the  spin resolved conductance. Indeed, we have
already become aware of two very recent related preprints
leading into these directions, one\cite{si} 
on Andreev reflection and spin injection into $s$- and
$d$-wave superconductors, and another\cite{enhance} discussing,
in conventional superconductors, 
out of equilibrium enhanced Andreev reflection with spin polarization.
\section{Acknowledgements}
We would like to thank J. Fabian, A.M. Goldman, A.J. Millis, S. Das Sarma, 
T. Venkatesan, V.A. Vas'ko, and S. Gasiorowicz for useful 
discussions. This work was supported by the US-ONR. 

%


\begin{references}
\bibitem{crit} These angles, for various scattering processes, are spin
dependent and imply the vanishing of the $x$ component of the 
appropriate wavevector, as used in Ref. {\protect{\onlinecite{zv}}}.
\bibitem{total} This is different from the case of total reflection where 
there is also  no contribution to the conductance above the energy gap.
\bibitem{sauls2} In Ref. {\protect{\onlinecite{sauls}}}, this is modeled by
an acceptance cone of angle $2\phi_c$ about the normal and with a uniform
weight function $1/2\phi_c$, where $\phi_c$ is barrier dependent.
\end{references}
\end{document}